\documentclass[a4paper,amsmath]{jpconf}
\usepackage{graphicx,epsfig}

\newcommand{\cmat}[2][ccccc]{\left( \begin{array}{#1} #2\\ \end{array}\right)}

\begin{document}

\title{Long-range interaction effects on neutrino oscillation}
\author{Hye-Sung Lee}
\address{Department of Physics, Brookhaven National Laboratory, Upton, NY 11973, USA}
\ead{hlee@bnl.gov}

\begin{abstract}
Motivated by the recent anomaly in the muon neutrino and anti-muon neutrino disappearance experiments, we consider a long-range interaction with an extremely light gauge boson and extraordinarily weak coupling.
A long-range interaction, consistent with current bounds, could have very pronounced effects on atmospheric neutrino disappearance that will be studied with the IceCube DeepCore array, currently in operation, and can have a significant effect on future high-precision long-baseline oscillation experiments. \\

\noindent
Talk given at NUFACT 11, XIIIth International Workshop on Neutrino Factories, Super beams and Beta beams, 1-6 August 2011, CERN and University of Geneva\footnote{\it This invited talk was based on the paper with Hooman Davoudiasl and William Marciano \cite{Davoudiasl:2011sz}.}
\end{abstract}

\section{Introduction}
MINOS is a long baseline neutrino oscillation experiment from the Fermilab to Soudan Mine in Minnesota.
It can study the muon neutrino and anti-muon neutrino oscillations over a baseline of $L = 735 ~{\rm km}$ with $E_\nu \sim {\rm GeV}$ scale.
The recent results of the MINOS $\nu_\mu$ and $\bar\nu_\mu$ disappearance experiments show disagreement \cite{Adamson:2011fa}.
This cannot be explained by the standard oscillation picture since $P(\nu_\mu \to \nu_\mu) = P(\bar\nu_\mu \to \bar\nu_\mu)$ in true vacuum by the CPT invariance, and the matter effect in the $\nu_\mu$ and $\bar\nu_\mu$ disappearance experiments at the MINOS are negligible.

It is likely due to the poor statistics especially in the $\bar\nu_\mu$ sector ($N_{\nu} \sim 2000$, $N_{\bar\nu} \sim 100$).
Nevertheless, it is still possible that this is a hint of new physics that can affect neutrino oscillations differently from the anti-neutrino oscillation.
In this talk, we go over a possible explanation and briefly discuss its implications for other neutrino experiments.
A detailed discussion can be found in Ref.~\cite{Davoudiasl:2011sz}.

\section{Long-range interaction and neutrino oscillation}
What type of new physics would be able to explain the MINOS anomaly?
If we look at the standard model interactions, the $W$ boson gives a potential only to the electron neutrinos, and it can affect neutrino oscillations involving the electron neutrino.
But the $Z$ boson gives a flavor-universal potential to the neutrinos making no effect in neutrino flavor oscillations:
$$H_{\rm SM} = V_W ~(1, 0, 0) + V_Z~ (1, 1, 1)$$
with $V_W = \sqrt{2} G_F n_e$, $V_Z = -G_F n_n / \sqrt{2}$ where $n_e$($n_n$) is the electron(neutron) number density.

We would like to consider a lepton flavor-dependent long-range interaction (LRI).
We assume a lepton flavor-dependent $U(1)$ gauge symmetry with an almost massless gauge boson $Z'$.
(Although this is our set up, it may not be necessary to consider an Abelian gauge symmetry or even a LRI to address the MINOS anomaly.)
$$H_{\rm LRI} = V_{Z'} \cmat{Q_e & & \\ & Q_\mu & \\ & & Q_\tau}$$
There are related works prior to our study about the LRI effects on neutrino oscillations and the MINOS anomaly explanation with a new interaction.
(For a limited list, see Refs.~\cite{Joshipura:2003jh,Grifols:2003gy,GonzalezGarcia:2006vp,Bandyopadhyay:2006uh,Engelhardt:2010dx,Mann:2010jz,Kopp:2010qt,Heeck:2010pg,Samanta:2010zh}.)

Under the new potential, the effective muon neutrino survival probability in the 2-flavor oscillation limit can be described by the effective mass splitting and mixing angle \cite{Joshipura:2003jh}: $\Delta \tilde{m}^2_{23} = \Delta m^2_{23} ([\xi - \cos(2\theta_{23})]^2 + \sin^2(2\theta_{23}) )^{1/2}$ and $\sin^2(2\tilde{\theta}_{23}) = \sin^2(2\theta_{23}) / ([\xi - \cos(2\theta_{23})]^2 + \sin^2(2\theta_{23}))$ with $\xi \equiv - 2 W_\tau E_\nu / \Delta m^2_{23}$ and $W_\tau = Q_\tau V_{Z'}$ (potential energy difference in $\nu_\mu - \nu_\tau$).
The $\xi$ flips sign for $\bar\nu$ causing different effects on $\nu$ and $\bar\nu$ unless $\sin^2(2\theta_{23}) = 1$.

In 1955, T.D. Lee and C.N. Yang discussed that LRIs get constraints from the E\"{o}tv\"{o}s-type experiments \cite{Lee:1955vk}.
The coupling constants (fine-structure constants) are constrained to be very small \cite{Dolgov:1999gk}: $\alpha' < 10^{-47}$ (for $Q =$ baryon number), $\alpha' < 10^{-49}$ (for $Q =$ lepton number).
So it is clear that we need astronomical sources to have any meaningful effects, which is possible since we are dealing with a LRI.
We assume $m_{Z'} < 1/AU \sim 10^{-18} ~{\rm eV}$ to include the Sun (but not much smaller to avoid galaxy and beyond).

For an anomaly-free $U(1)'$ charges, we choose $Q = (B-L) + (L_\mu - L_\tau) = B - L_e -2 L_\tau$ or $H_{\rm LRI} = V_{Z'} ~(-1, 0, -2)$.
Neutrons in the Sun and the Earth are the source of the new potential (due to $B-L$), and the flavor-dependent charges ($L_\mu - L_\tau$) affect $\nu$ flavor oscillation.

The Sun and the Earth have neutrons of $N_n^\odot = 1.7 \times 10^{56}$ and $N_n^\oplus = 1.8 \times 10^{51}$, respectively, and the new potential by the neutrons is given by $V_{Z'} \sim (\alpha' / 10^{-50}) \times {\cal O}(10^{-12} ~\rm{eV})$.
Since the MINOS $\nu$ oscillation is relevant to $\Delta m_{23}^2 / E_\nu \sim {\cal O}(10^{-12} ~\rm{eV})$, a LRI with $\alpha' \sim {\cal O} (10^{-50})$ level can affect MINOS experiments significantly.
In other words, the $\nu$ oscillation experiments are a good probe of an extremely weak LRI.

We performed a rough fit to the MINOS data (for both the neutrinos and anti-neutrinos) with the LRI, and found the best-fit point at $\Delta m_{23}^2 = 2.4 \times 10^{-3} ~{\rm eV}^2$, $\sin^2 (2\theta_{23}) = 0.9$, $\alpha' = 1.0 \times 10^{-52}$ (or $W_\tau = 5.6 \times 10^{-14} ~{\rm eV}$), using the simplified MINOS data from Ref.~\cite{Kopp:2010qt}.
(The MINOS data constrains $\alpha' < 5 \times 10^{-52}$ at roughly $3\sigma$ level.)

This best-fit point is not positively ruled out by solar+KamLAND $\nu$ and atmospheric $\nu$ data, although there is a tension between the potential for MINOS data and SK atmospheric $\nu$ data (See \ref{sec:appendix}).
This best-fit with LRI does not really improve the goodness-of-fit of the MINOS data over the SM fitting.
We rather take it as a motivated benchmark point to explore other experiments to test the LRI idea.
For this purpose, we want to consider two setups, the IceCube DeepCore experiment, and the future long baseline neutrino experiment (LBNE).

\section{Implications for IceCube DeepCore and future LBNE experiments}
DeepCore is composed of 6 additional densely instrumented strings plus 7 nearby IceCube strings, and it is located in the bottom center of the IceCube experiment at the South Pole.
It was recently commissioned and the analysis of the first year data is in progress (see F. Halzen's talk).
DeepCore is a high statistics detector that can study the atmospheric neutrino with precision \cite{FernandezMartinez:2010am}.
It can trigger ${\cal O}(10^5)$ atmospheric neutrinos per year in $E_\nu \approx 1-100 ~\rm{GeV}$, which is complementary to the original IceCube (optimized for $E_\nu > 10 ~{\rm TeV}$).

\begin{figure}[t]
\begin{minipage}{17pc}
\includegraphics[height=10pc]{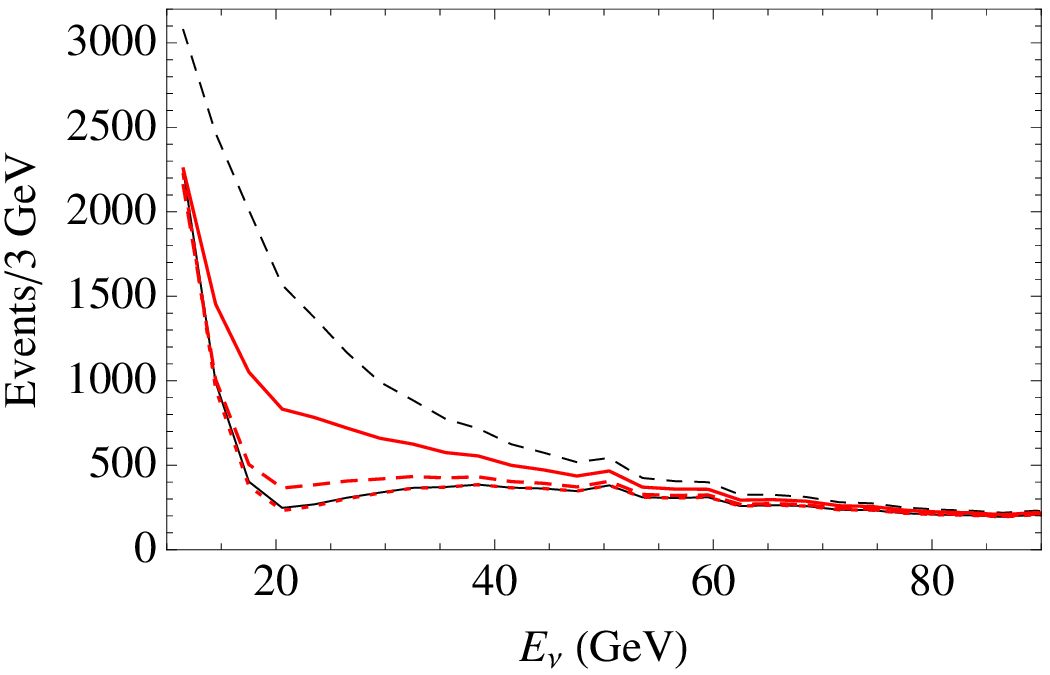}
\caption{\label{fig:deepcore} Atmospheric muon neutrino disappearance at DeepCore. The curves are the unoscillated oscillation (black dashed), standard oscillation (black solid), and the MINOS best-fit (red solid).}
\end{minipage} \hspace{2pc}%
\begin{minipage}{17pc}
\includegraphics[height=10pc]{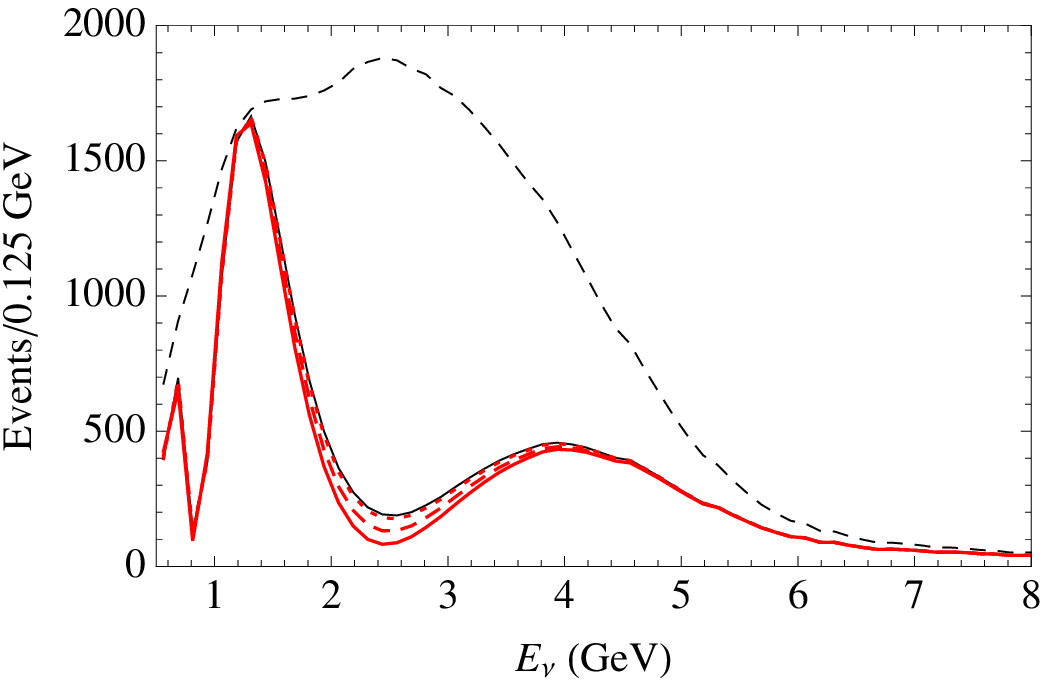}
\caption{\label{fig:dusel} The $\nu_\mu$ disappearance at the future LBNE ($L = 1300 ~{\rm km}$, $200 ~{\rm kton}$ water Cherenkov detector, 5-years run). The shapes and colors of the curves are the same as Fig.~\ref{fig:deepcore}.}
\end{minipage}
\end{figure}

We would like to see the LRI effects on the atmospheric muon neutrino disappearance at DeepCore.
Figure~\ref{fig:deepcore} shows the numerical results without a full detector simulation. 
The black dashed curve is for the unoscillated case, the black solid curve is for the standard oscillation, and the red solid curve is for the MINOS best-fit case.
It shows that LRI effects on the DeepCore atmospheric muon neutrino disappearance experiment are quite distinct from the standard oscillation.

Another interesting aspect that we may be able to observe at the DeepCore experiment is the annual modulation in the new potential.
Since the distance between the Sun and the Earth changes over the year (about 3\% level), the new potential would have an annual modulation, which means that the effective oscillation parameters might change with seasons.
With an assumption of uniform flux (before oscillation) over the year, our calculation shows it can give a percent level annual modulation in the muon neutrino disappearance experiment, with a total number of a few thousands per year.

Now, we consider the implication for the future LBNE.
Figure~\ref{fig:dusel} shows the muon neutrino disappearance experiments for the envisioned LBNE at DUSEL.
It assumes the baseline of $L = 1300 ~{\rm km}$ and 200 kton water Cherenkov detector \cite{DUSEL}.
As it shows, for the first minimum, the muon neutrino with the new potential has fewer events than the standard oscillation, which means larger $\sin^2 (2\theta_{23})$.
The corresponding result for the muon antineutrino has more events (or smaller $\sin^2 (2\theta_{23})$) at the first minimum.
The LBNE can tell the different LRI effects on $\nu$ and $\bar\nu$.
The observation of the annual modulation effect, however, may need enhanced capabilities compared to what we considered.

\section{Implications for the charged lepton sector}
Let us make a brief comment on the implications of the LRI for the charged lepton sector.
We have a lepton-flavor-dependent $U(1)$ gauge symmetry with a nearly massless gauge boson $Z'$.
When we have a flavor-dependent U(1) gauge symmetry, $Z'$-mediated flavor changing neutral currents (FCNCs) at tree-level are present, in general.
For instance, $\mu$-$e$-$Z'$ vertex may exist in the mass eigenstate.

The question is whether we can neglect this $Z'$-mediated FCNC because it has a tiny coupling ($\alpha' < 10^{-50}$) as constrained by the E\"{o}tv\"{o}s-type experiments.
We cannot neglect this, in general, because of the Goldstone boson equivalence principle.
For instance, the $\mu \to e Z'$ decay is enhanced since $m_{Z'} \ll m_\mu$.
Then, $\Gamma(\mu \to e Z') \sim \alpha' m_\mu (m_\mu^2 / m_{Z'}^2)$ can lead to a too fast muon decay because of the extremely small $m_{Z'}$.
Thus, explicit model buildings with Higgs sector should address this issue.

\section{Summary}
MINOS data may be a hint for the new physics which acts differently on neutrinos and antineutrinos.
The lepton-flavor-dependent LRI with $\alpha' \sim 10^{-52}$ is such a possibility since neutrino oscillations are sensitive to it.
The IceCube DeepCore experiment, which is an ongoing experiment, can test this possibility.
Especially, if the annual modulation is observed, it can point to the solar origin of the new potential.
A future LBNE can test if the LRI gives different effects on the neutrino and antineutrino oscillations.
LRI effects on the charged lepton sector are not negligible, in general, and depend on the explicit Higgs sector model building.

\ack{
This work was supported in part by the U.S. Department of Energy under Grant Contract No. DE-AC02-98CH10886.
It is a great pleasure to thank my collaborators H. Davoudiasl and W. Marciano.
}

\appendix
\section{Constraints on the new potential}
\label{sec:appendix}
Our MINOS best-fit is at $\alpha' = 1.0 \times 10^{-52}$ or at the potential energy $W_\tau = 5.6 \times 10^{-14} ~{\rm eV}$.

The constraint on the LRI from the solar neutrino and KamLAND reactor neutrino was studied in Refs.~\cite{GonzalezGarcia:2006vp,Bandyopadhyay:2006uh}, which is, for our model, $\alpha' < 5 \times 10^{-52}$ at $3\sigma$ level.

The constraints from the atmospheric neutrino data measured at the Super-Kamiokande is given in terms of the $\epsilon_{\tau\tau}$ with $\epsilon_{\tau\tau} < 0.2$ at $95\%$ CL \cite{Friedland:2004ah}.
Its corresponding value for the potential energy ($W_\tau = \epsilon_{\tau\tau} \sqrt{2} G_F n_e$) depends on $n_e$ (electron number density).
The Earth core ($R < 3400 ~{\rm km}$) has $n_e \approx 12 ~{\rm g/cm}^3$ ~($W_\tau < 9 \times 10^{-14} ~{\rm eV}$) and the mantle ($R > 3400 ~{\rm km}$) has $n_e \approx 5 ~{\rm g/cm}^3$ ~($W_\tau < 4 \times 10^{-14} ~{\rm eV}$).
Our MINOS best-fit potential energy is somewhere in the middle, which shows some tension though it calls for a more detailed analysis in order to pinpoint a precise constraint on the LRI.

\section*{References}

\end{document}